\documentclass[prd,twocolumn,showpacs,amsmath,amssymb]{revtex4}
\usepackage{bm}

\begin{document}

%%%%%%%%%%%%%%%%%%%%%%%%%%%%%%%%%%%%%%%%%%%%%%%%%%%

\title{Geometric interpretation of spinor and gauge vector}
\author{Haigang Lu}
\email{lhg@chem.pku.edu.cn}
\affiliation{Quantum chemistry group, Department of chemistry,
Peking University, Beijing 100871, China}

\date{\today}

%%%%%%%%%%%%%%%%

\begin{abstract}
In the standard model, electroweak theory is based on chiral gauge symmetry,
which only contains massless fermions and gauge vectors. 
So the fundamental field equations are Weyl equations including
gauge vectors.
We deduced these Weyl equations without gauge vector from quadric surface 
in three dimensional oriented projective geometry.
It showed that Weyl equations were equations of generating lines in the quadric
surface.
In other word, the geometric correspondence of spinor is generating lines of
quadric surface in momentum space.
It also showed that two spinors would generate a gauge 
vector.
After this quadric surface was localized, its off-origin generating 
line would correspond to global Weyl equation including gauge vector, 
from which we deduced the global and local gauge transformation.
\end{abstract}


\maketitle

%%%%%%%%%%%%%%%%%%%%%%%%%%%%%%%%%%%%%%%%%%%%%%%%%%%%%%%%%%%%

\section{Introduction}

In modern physics, spinor analysis has been a basis in field theory, 
such as unified field theory, general relativity.
It is well known that spinor is more fundamental than vector, 
because a vector is a dyad of conjugate spinors.
Though spinor is so fundamental, we can not visualize it in our mind.
The simplest, two-component spinor comes from Weyl equation, 
i.e. field equation of massless fermions. 
Then, we expect to be able to find the geometric correspondence of spinors 
by the geometric interpretation of Weyl equations.

This paper includes five parts:
First, Weyl equations are reviewed briefly in \ref{weyl}.
Secondly, the special relativity are described by a convenient geometric
method in \ref{proj}.
Thirdly, the equivalence between equations of generating lines in quadric 
surface and Weyl equations is deduced in \ref{geom}.
Fourthly, the geometric interpretation of spinor representation of gauge vector
are given in \ref{vect}. 
And in \ref{local} we deduce the global and local gauge transformation.
At the end of paper, we conclude in \ref{conc}.

We use a system of unit in which $\hslash=c=1$ in this paper.

%%%%%%%%%%%%%%%%%%%%%%%%%%%%%%%%%%%%%%%%%%%%%%

\section{Weyl equations}
\label{weyl}

%%%%%%%%%%%%%%%%%%
Historically, Dirac\cite{dirac} formulated his four-component relativistic wave
equation in 1928.
Subsequently, Weyl\cite{weyl} proposed two-component relativistic wave equation 
to describe the massless fermions in 1929.
But it was used as neutrino field equation until 1957\cite{salam, LeeYang,%
Landau}.
Later, the maximum parity violation of weak interaction made Weyl
equation become the fundamental fermion field equations in electroweak 
theory\cite{wein, salam1},
because this two-component equation satisfies the chiral gauge 
symmetry of electroweak theory as required.

Weyl equations have two types, i.e.
\begin{subequations}
\begin{eqnarray}
\partial_t\psi&=&\bm{\sigma}\cdot\nabla\psi\label{weyle1}\\
\noalign{and}
\partial_t\tilde{\psi}&=&-\bm{\sigma}\cdot\nabla\tilde{\psi}\label{weyle2},
\end{eqnarray}
\end{subequations}
where the $\sigma$'s are $2\times 2$ matrices.
One of suitable choice of this matrices
is Pauli spin ones:
\begin{equation}
\sigma_1=\left(\begin{array}{cc}0&1\\1&0\end{array}\right),\quad
\sigma_2=\left(\begin{array}{cc}0&-i\\i&0\end{array}\right),\quad
\sigma_3=\left(\begin{array}{cc}1&0\\0&-1\end{array}\right).
\end{equation}
The wavefunction $\psi$ must have two components, since it is operated on
by $2\times 2$ matrices. 
Thus,
\begin{equation}
\psi=\left(\begin{array}{c}\psi_1\\\psi_2\end{array}\right).
\end{equation}
Then, Weyl equations satisfy the relativistic relation:
\begin{equation}
E^2=\bm{p}^2.
\end{equation}

In fact, the equation (\ref{weyle1}) and (\ref{weyle2}) are correlative, 
because if $\psi$ is a solution of equation (\ref{weyle1}), 
it is easily proved that
\begin{equation}\label{corr}
\tilde{\psi}\equiv\sigma_2\psi^*
\end{equation}
is a solution of equation (\ref{weyle2}),
where $\psi^*$ is complex conjugate of $\psi$.

Consider the plane-wave solutions of Weyl equation (\ref{weyle1}):
\begin{equation}\label{moment}
\psi(\bm{x},t)=\psi(\bm{p},E)e^{i(\bm{px}-Et)}.
\end{equation}
Substituting (\ref{moment}) in (\ref{weyle1}) we get
\begin{equation}
\bm{\sigma}\cdot \bm{p}\psi(\bm{p},E)=-E\psi(\bm{p},E),
\end{equation}
i.e.
\begin{equation}
\left(\begin{array}{rcrcrcr}
p_z&+&E&&p_x&-&ip_y\\p_x&+&ip_y&&-p_z&+&E
\end{array}
\right)\left(\begin{array}{c}
\psi_1\\\psi_2\end{array}\right)=0.
\end{equation}
In the same way, we can get the following equation from (\ref{weyle2}):
\begin{equation}
\bm{\sigma}\cdot \bm{p}\tilde\psi(\bm{p},E)=E\tilde\psi(\bm{p},E).
\end{equation}
After defining helicity operator to be $H=\bm{\sigma}\cdot\bm{p}/|\bm{p}|$,
it is deduced that in Weyl equation (\ref{weyle1}) the eigenvalue of
$H$ is always $-1$ for positive energy and $+1$ for
negative one, i.e., positive particles are in left-handness and negative
one in right-handness.
On the contrary, in (\ref{weyle2}) positive particles are in right-handness
and negative one in left-handness.
On the other hand,
we can get Dirac massless equation from one first type and one second type
Weyl equation, or from two first type ones because of correlativity 
(\ref{corr}).
%%%%%%%%%%%%%%%%%%%%%%%%%%%%%%%%%%%%%%%%%%%%%%%

\section{Geometry}
\label{proj}

%%%%%%%%%%%%%%%%%%

Now we shall think of the minimal homogeneous (Klein) geometry which Weyl
equations could be able to be embedded.
We call this geometry the null relativistic geometry for without mass term.

First, we could see that either of Weyl equations have four terms, i.e.,
\begin{equation}\label{weyla}
\sigma_1p_x\psi+\sigma_2p_y\psi+\sigma_3p_z\psi\pm E\psi=0.
\end{equation}
This requires that there must be at least four coordinates in this homogeneous 
geometry.

Second, the components of wave function in Weyl equations (\ref{weyla})
are all complex functions, then the geometry should be complex.
Then the required homogeneous geometry is of four complex coordinates.

Third, according to Klein's Erlangen program \cite{Klein}, 
objects of geometry are the invariant and invariance under transformation group, 
then we shall determine the transformation group of this geometry,
which includes the symmetry group of Weyl equations as a subgroup.

As the relativistic field equations of fermions,
Weyl equations satisfy naturally the requires of relativity.
In contrast with special and general relativity in table below,
Weyl equations have more symmetry than special relativity.
Its symmetry group is conformal group, $SO(4,2)\simeq SU(2,2)$.
So, this requires that the transformation group of this geometry must include 
$SU(2,2)$ as a subgroup.
\begin{widetext}
\begin{center}
\begin{tabular}{|r||c|c|c|}
\hline
Relativity&Null&Special&General \\\hline\hline
Mass&&&gravitational\\
term&no&inertial&($\cong$ inertial)\\\hline
Symmetry&conformal group &Poincar\'e group&Lorentz group\\
group&$SO(4,2)\simeq SU(2,2)$&$SO(3,1)\otimes P(4)$&$SO(3,1)$\\\hline
Constraint&$\bm{p}^2=E^2$&$\bm{p}^2+m^2=E^2$&
$G_{\mu\nu}=-8\pi G T_{\mu\nu}$\\\hline
Geodesic&$0=\bm{x}^2-t^2$&$ds^2=\bm{x}^2-t^2$&
$ds^2=g_{\mu\nu}dx^\mu dx^\nu$\\\hline
\end{tabular}
\end{center}
\end{widetext}

All in all, the null relativistic geometry is complex geometry,
which has four coordinates, and whose transformation group includes $SU(2,2)$ 
as a subgroup.
Then the minimal geometry is three dimensional oriented complex projective 
geometry $O\mathbb{C}P^3=\mathbb{C}P^3\otimes Z_2$ \cite{Stolfi,sam},
whose transformation group is $SL(4,\mathbb{C})$, 
or four dimensional complex vector space $\mathbb{C}^4$ \cite{GW},
whose transformation group is $GL(4,\mathbb{C})=SL(4,\mathbb{C})\otimes 
U(1,\mathbb{C})$.
The projective geometry is preferred to discuss the geometric mean of 
Weyl equations.
To express a point in three dimensional projective geometry,
we need four homogeneous coordinates $(x_1,x_2,x_3,x_4)$,
In this way, the ordinary points could be expressible as $(x_1,x_2,x_3,1)$,
and infinity ones as $(x_1,x_2,x_3,0)$.
While considering the geometric expression in four dimensional spacetime,
we shall turn to $\mathbb{C}^4$.

%%%%%%%%%%%%%%%%%%%%%%%%%%%%%%%%%%%%%%%%%%%%%%%

\section{Geometric interpretation}
\label{geom}

%%%%%%%%%%%%%%%%%%

Now we shall find the geometric interpretations of Weyl equations in 
$O\mathbb{C}P^3$.
The goal is geometric correspondence of Weyl equations.

Because Weyl equations must satisfy the constrain $p_x^2+p_y^2+p_z^2-E^2=0$,
its geometric correspondence would lie in the quadratic surfaces 
$x_1^2+x_2^2+x_3^2+x_4^2=0$ of $O\mathbb{C}P^3$.
Now, we decompose this quadratic surface as below:
\begin{eqnarray*}
&x_1^2+x_2^2+x_3^2+x_4^2&=0\nonumber\\
\Rightarrow&(x_1+ix_2)(x_1-ix_2)&=-(x_3+ix_4)(x_3-ix_4)
\end{eqnarray*}
so, 
\begin{subequations}\label{line}
\begin{eqnarray}
-\frac{\displaystyle x_1-ix_2}{\displaystyle x_3+ix_4}
=\frac{\displaystyle x_3-ix_4}{\displaystyle x_1+ix_2}
&=&\frac{\psi_1}{\psi_2}=\lambda\label{linea}\\
-\frac{\displaystyle x_1-ix_2}{\displaystyle x_3-ix_4}
=\frac{\displaystyle x_3+ix_4}{\displaystyle x_1+ix_2}
&=&\frac{\tilde\psi_1}{\tilde\psi_2}=\mu.\label{lineb}
\end{eqnarray}
\end{subequations}
In expressions of ${\psi_1}/{\psi_2}$ and ${\tilde\psi_1}/{\tilde\psi_2}$,
$(\psi_1,\psi_2)$ and $(\tilde\psi_1,\tilde\psi_2)$ are homogeneous
complex coordinates of parameters $\lambda$ and $\mu$.
Each of equations \eqref{line} is a  generating line(or generator)
of quadric surface $x_1^2+x_2^2+x_3^2+x_4^2=0$.

After substituting $x_1$, $x_2$, $x_3$, and $ix_4$ with $p_x$, $p_y$,
$p_z$, and $E$ respectively in (\ref{line}), 
we get
\begin{subequations}
\begin{eqnarray}
-\frac{p_x-ip_y}{p_z+E}=\frac{p_z-E}{p_x+ip_y}&=&\frac{\psi_1}{\psi_2}\\
-\frac{p_x-ip_y}{p_z-E}=\frac{p_z+E}{p_x+ip_y}&=&
\frac{\tilde\psi_1}{\tilde\psi_2}
\end{eqnarray}
\end{subequations}
After writing as matrix, they are just Weyl equations:
\begin{subequations}\label{weyl0}
\begin{eqnarray}
\left(\begin{array}{cc}p_z+E&p_x-ip_y\\p_x+ip_y&-p_z+E\end{array}\right)
\left(\begin{array}{c}\psi_1\\\psi_2\end{array}\right)&=&0\\
\left(\begin{array}{cc}p_z-E&p_x-ip_y\\p_x+ip_y&-p_z-E\end{array}\right)
\left(\begin{array}{c}\tilde\psi_1\\\tilde\psi_2\end{array}\right)&=&0 
\end{eqnarray}
\end{subequations}
From here, we can see that $\tilde\psi=(\tilde\psi_1,\tilde\psi_2)^T=
\sigma_2\psi^*$, where $\psi=(\psi_1,\psi_2)^T$, and superscript $T$
is transposition of matrix.
Then we see that Weyl equations are two families of generating lines of
constrained quadric $\bm{p}^2-E^2=0$ in momentum-energy space,
and the corresponding two-component spinors $(\psi_1,\psi_2)$ and 
$(\tilde\psi_1,\tilde\psi_2)$ are homogeneous coordinates of these two families
of generating lines in this quadric surface.
So the correspondence between physics and geometry is
\begin{center}
\begin{tabular}{|rcl|}\hline
Physical object&$\leftrightarrow$&Geometric object\\\hline\hline
Weyl equation&$\leftrightarrow$&line in quadric surface\\\hline
two-component spinor&$\leftrightarrow$&homogeneous coordinates\\
&& of the above line\\\hline
\end{tabular}
\end{center}

Now, let us see where the gauge degree of freedom come from.
As we see in \eqref{line}, the $\psi_1/\psi_2$ will not change as $\psi_1$ and 
$\psi_2$ are transform to $\psi_1\cdot C_0$ and $\psi_2\cdot C_0$,
where $C_0=R\cdot\exp(i\alpha)$ may be a complex number or function 
of coordinates.
Because $R$ will be absorb in the normalization constant of wavefunction, 
there is only one non-physical degree of freedom for this parameter 
representation, i.e. $\exp(i\alpha)$,
that is the gauge transformation 
$$\psi\to\psi\exp(i\alpha)$$
of wavefunction of massless fermion.

Based on the above geometric interpretation, we can deduce 
the internal symmetry of fermions.
In $O\mathbb{C}P^3$, line space is Grassmann space $G(2,4)\otimes Z_2=
S^2\times S^2$,
which has four dimensions, then spinors of fermions have four degrees
of freedom, and their compact simple symmetry group is at most $SU(4)$, 
the maximum real subgroup of transformation group $SL(4,C)$ of $O\mathbb{C}P^3$.

%%%%%%%%%%%%%%%%%%%%%%%%%%%%%%%%%%%%%%%%%%%%%%%

\section{Gauge vectors}
\label{vect}

%%%%%%%%%%%%%%%%%%

After finding the geometric interpretation of spinor, 
we will discuss how to express the gauge vectors by these spinors,
because only gauge vector bosons are same fundamental as fermions 
in standard model.

A quadric surface is the union of two families of generating lines.
From equations \eqref{line}, we can get the expression of this quadric surface
by the two parameters $\lambda$ and $\mu$ of generating line, i.e.,
\begin{equation*}
x_1:x_2:x_3:x_4=i(1-\mu\lambda):(1+\mu\lambda):i(\mu+\lambda):(\mu-\lambda).
\end{equation*}
After replaced $\lambda, \mu$ by $\psi_1/\psi_2, \psi'_1/\psi'_2$ and 
$x_1, x_2, x_3, x_4$ by $A_1, A_2, A_3, iA_4$, we get
\begin{equation}\label{gauge}
\left(\begin{array}{c}
A_1^0 \\
A_2^0 \\
A_3^0 \\
A_4^0
\end{array}\right)=\left(\begin{array}{cccc}
-1 &0  &0  &1  \\
-i &0  &0  &-i  \\
 0&  1&  1& 0 \\
 0&  1&  -1& 0
\end{array}\right)\left(\begin{array}{c}
\psi_1{\tilde\psi}'_1\\
\psi_1{\tilde\psi}'_2\\
\psi_2{\tilde\psi}'_1 \\
\psi_2{\tilde\psi}'_2
\end{array}\right).
\end{equation}
Because the $\lambda$ and $\mu$ are two variables, so this vector has two
degrees of freedom, just as required by gauge vector.
We consider this intersection point in quadric surface of three dimensional 
projective geometry as the geometric correspondence of gauge vector,
such as photon.

At the end of section \ref{geom} we found fermions had only four complex 
dimensions and their internal symmetry is $SU(4)$ in $O\mathbb{C}P^3$.
Because gauge bosons is dyads of two of these spinors, 
they would correspond with the adjoint representation of fermions.
Then the transformation of gauge boson is determined by gauge transformation
of fermions.
And according to \eqref{gauge}, the corresponding gauge vector 
would transform as below:
$$
A^0_\mu\to A^0_\mu\exp(i\alpha)\exp(-i\tilde\alpha').
$$

%%%%%%%%%%%%%%%%%%%%%%%%%%%%%%%%%%%%%%%%%

\section{local gauge transformation}
\label{local}

%%%%%%%%%%%%%%%%%%%%%%%

Now, we will discuss the local transformation of Weyl equation in 
four dimensional complex vector space $\mathbb{C}^4$ instead of
three dimensional oriented projective geometry $O\mathbb{C}P^3$.
The constraint $\bm{p}^2-E^2=0$ is a quadric supersurface in $\mathbb{C}^4$,
whose origin is in the point $(0,0,0,0)$.
For a general quadric supersurface, whose origin is in $(\bm{p}_0, E_0)$,
this constraint is $(\bm{p}-\bm{p}_0)^2-(E-E_0)^2=0$.
This is just the local transformation in homogeneous space.

If we combine this local transformation with the gauge one,
we will find, the total transformation is 
$$
\begin{array}{rrl}
(\bm{p},E)&\xrightarrow{\mathrm{localization}}&(\bm{p}-\bm{p}_0, E-E_0)\\
&\xrightarrow{\mathrm{gauge}}& (\bm{p}-\bm{p}_0, E-E_0)
\exp(i\alpha)\exp(-i\tilde\alpha'),
\end{array}
$$
for $(\bm{p}-\bm{p}_0, E-E_0)$ also satisfies the equations \eqref{gauge}, 
which transforms as the complex parameters of line,

From above analysis, we will deduce the gauge transformation of gauge
vector in Weyl equations.
At first, let
$$
p_\mu = (\bm{p},E).
$$
we get
$$
\sigma_\mu p_\mu\psi=0,
$$
where the $\sigma_\mu$ is Pauli matrix and identity $2\times 2$ matrix.
After generalizing to the off-origin locality, we get
$$
\to\;\sigma_\mu (p_\mu-A_\mu)\psi=0.
$$
we could insert a identity transformation matrix $I=\exp(-i\alpha)\exp(i\alpha)$
in front of $\psi$ and keep the equation unchanged.
Then
$$
\to\;\sigma_\mu (p_\mu-A_\mu)\cdot\exp(-i\alpha)\exp(i\alpha)\psi=0.
$$
Then, we get the global fundamental gauge transformation:
$$
\to\;\sigma_\mu (p_\mu\exp(-i\alpha)-A_\mu\exp(-i\alpha))\exp(i\alpha)\psi=0.
$$

To get the covariant differential, we will focus on the evolvement
$$p_\mu\to\partial_\mu\to D_\mu.$$
We take the quantization
$$p_\mu\to\; \partial_\mu=-i\hslash\partial/\partial x_\mu,$$ 
and the above expression become
$$
\to \;\partial_\mu\exp(-i\alpha)-eA_\mu\exp(-i\alpha),
$$
where $e$ is a coupling constant, such as electronic charge.
After expanding the exponential function to second order of $\alpha$ 
and/or $e$, we get
$$
\to \; \partial_\mu+\partial_\mu(-i\alpha) +\partial_\mu(-i\alpha)^2/2
-ieA_\mu-eA_\mu(-i\alpha).
$$
And deduced to first order of $\alpha$ and/or $e$:
$$
\to\; \partial_\mu+\partial_\mu (-i\alpha) -ieA_\mu,
$$
i.e, the standard covariant differential of gauge principle:
$$
\to \;\partial_\mu-ie(A_\mu+\frac{1}{e}\partial_\mu\alpha)
$$

%%%%%%%%%%%%%%%%%%%%%%%%%%%%%%%%%%%%%%%%%%%%%%%

\section{conclusion}
\label{conc}

%%%%%%%%%%%%%%%%

Now, we conclude below:

The Weyl equations correspond with equations of generating line of
quadric surface in three dimensional oriented projective geometry,
and spinor is just the complex parameter coordinate of above line.
In these two families of generating lines, every line in one family
always intersect with one in another family, then the point of intersection
just corresponds with the gauge vector.

The homogeneous parameter coordinate of line allow a non-physical,
gauge transformation $\exp(i\alpha)$, which lead to the standard
covariant transformation of gauge vector in Weyl equations with
interaction.
So, the local gauge transformation of fermion and gauge vector
is not independent.
We also get the global gauge fundamental transformation.

%%%%%%%%%%%%%%%%%%%%%%%%%%%%%%%%%%%%%%%%%%%%

%%%%%%%%%%%%%%%%%%%%%%%%%%%%%%%%%%%%%%%%%


\begin{thebibliography}{99}

\bibitem{dirac} P.A. Dirac, \textit{Proc. Roy. Soc.}, \textbf{A117}, 610(1928);
\textbf{A118}, 351(1928).

\bibitem{weyl} H. Weyl, \textit{Zeitschrift f\"ur physik}, \textbf{56},
330(1929).

\bibitem{salam} A. Salam, \textit{Il Nuovo cimento}, \textbf{5}, 299(1957).

\bibitem{LeeYang} T.D. Lee and C.N. Yang, \textit{Physical Review},
\textbf{105}, 1671(1957).

\bibitem{Landau} L. Landau, \textit{Nuclear Physics}, \textbf{3}, 127(1957).

\bibitem{wein} S. Weinberg, Phys. Rev. Lett. \textbf{19}, 1264 (1967).

\bibitem{salam1} A. Salam, in \textit{Elementary Particle Theory}, 
edited by N. Svartholm (Almqvist and Wiksell, Stockholm, 1969), p.367.

\bibitem{Klein} F. Klein, \textit{Gesammelte Mathematische Abhandlungen,
Erster Band} (Springer-Verlag, Berlin, 1973), p. 460.

\bibitem{Stolfi} J. Stolfi, \textit{Oriented Projective Geometry}, 
(Academic Press, 1991).

\bibitem{sam} P. Samuel, \textit{Projective Geometry},
(Springer-Verlag, Heidelberg, 1988).

\bibitem{GW} K.W. Gruenberg and A.J. Weir, \textit{Linear Geometry},
(GTM, Springer-Verlag, New York, 1977). 

\end{thebibliography}
\end{document}